# The Open Access Citation Advantage: Quality Advantage Or Quality Bias?

## Open Access Archivangelism

### Sunday, January 21. 2007

#### The Open Access Citation Advantage: Quality Advantage Or Quality Bias?

**SUMMARY:** *Many studies have now reported the positive correlation between Open Access (OA) self-archiving and citation counts ("OA Advantage," OAA). But does this OAA occur because (QB) authors are more likely to self-selectively self-archive articles that are more likely to be cited (self-selection "Quality Bias": QB)? or because (QA) articles that are self-archived are more likely to be cited ("Quality Advantage": QA)? The probable answer is both. Three studies [by (i) Kurtz and co-workers in astrophysics, (ii) Moed in condensed matter physics, and (iii) Davis & Fromerth in mathematics] had reported the OAA to be due to QB [plus Early Advantage, EA, from self-archiving the preprint before publication, in (i) and (ii)] rather than QA. These three fields, however, (1) have less of a postprint access problem than most other fields and (i) and (ii) also happen to be among the minority of fields that (2) make heavy use of prepublication preprints. Chawki Hajjem has now analyzed preliminary evidence based on over 100,000 articles from multiple fields, comparing* self-selected *self-archiving with* mandated *self-archiving to estimate the contributions of QB and QA to the OAA. Both factors contribute, and the contribution of QA is greater.*

This is a preview of some preliminary data (not yet refereed), collected by my doctoral student at UQaM, Chawki Hajjem. This study was done in part by way of response to Henk Moed's replies to my comments on Moed's (self-archived) preprint:

> Moed, H. F. (2006) The effect of 'Open Access' upon citation impact: An analysis of ArXiv's Condensed Matter Section

Moed's study is about the "Open Access Advantage" (OAA) -- the higher citation counts of self-archived articles -- observable across disciplines as well as across years (from Hajjem et al. 2005; red bars are the OAA):

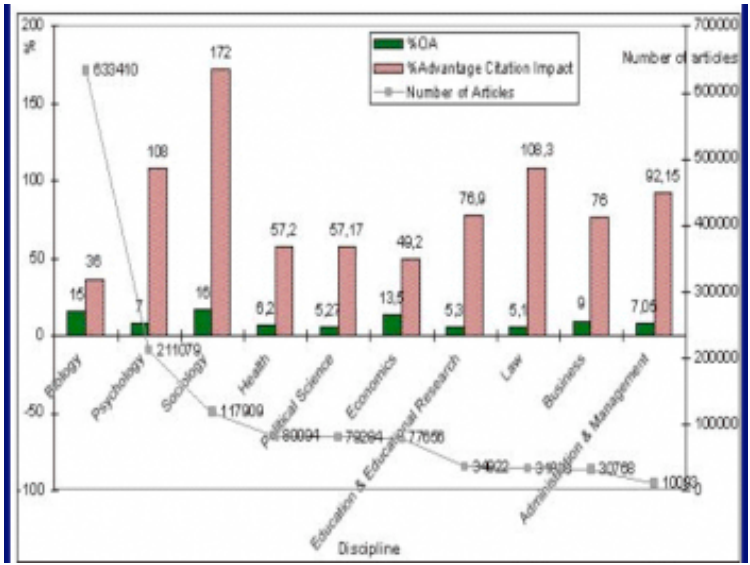

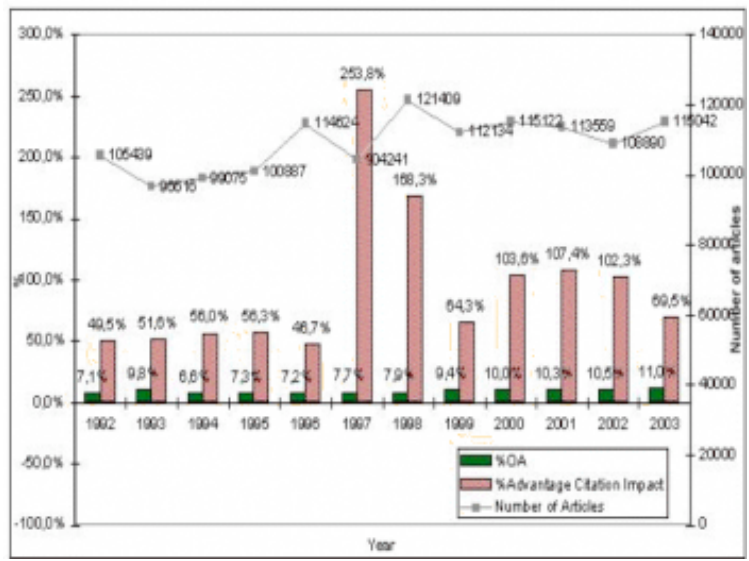

**FIGURE 1. Open Access Citation Advantage By Discipline and By Year.**
Green bars are percentage of articles self-archived (%OA); red bars, percentage citation advantage (%OAA) for self-archived articles for 10 disciplines (upper chart) across 12 years (lower chart, 1992-2003). Gray curve indicates total articles by discipline and year.
**Source**: Hajjem, C., Harnad, S. and Gingras, Y. (2005) Ten-Year Cross-Disciplinary Comparison of the Growth of Open Access and How it Increases Research Citation Impact. *IEEE Data Engineering Bulletin* 28(4) pp. 39-47.

The focus of the present discussion is the factors underlying the OAA. There are at least five potential contributing factors, but only three of them are under consideration here: **(1)** Early Advantage (**EA**), **(2)** Quality Advantage (**QA**) and **(3)** Quality Bias (**QB** -- also called "Self-Selection Bias").

**OA Advantage**

**OAA = EA + QA + UA + (CA) + (QB)**

- **EA: Early Advantage:** Self-archiving preprints before publication increases citations (higher-quality articles benefit more)
- **QA: Quality Advantage:** Self-archiving postprints upon publication increases citations (higher-quality articles benefit more)
- **UA: Usage Advantage:** Self-archiving increases downloads (higher-quality articles benefit more)
- **(CA: Competitive Advantage):** OA/non-OA advantage (CA disappears at 100%OA)
- **(QB: Quality Bias):** Higher-quality articles are self-selectively self-archived more (QB disappears at 100%OA)

Preprints that are self-archived before publication have an Early Advantage (EA): they get read, used and cited earlier. This is uncontested.

> Kurtz, Michael and Brody, Tim (2006) The impact loss to authors and research. In, Jacobs, Neil (ed.) *Open Access: Key strategic, technical and economic aspects*. Oxford, UK, Chandos Publishing.

In addition, the proportion of articles self-archived at or after publication is higher in the higher "citation brackets": the more highly cited articles are also more likely to be the self-archived articles.

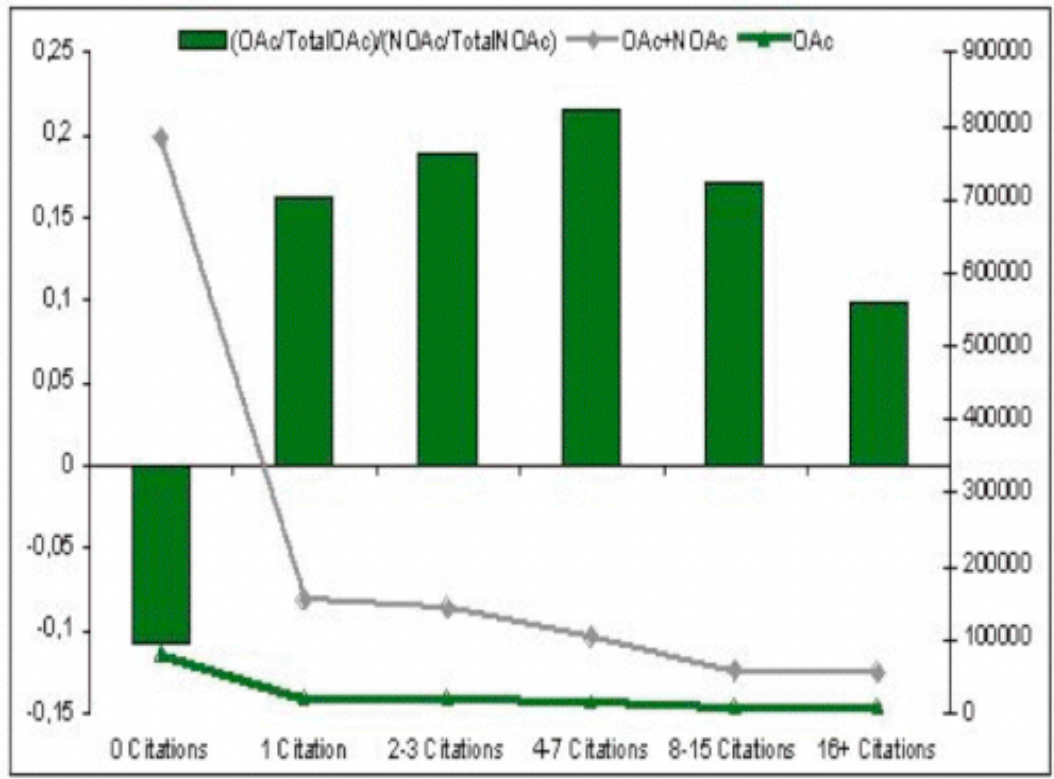

**FIGURE 2. Correlation between Citedness and Ratio of Open Access (OA) to Non-Open Access (NOA) Ratios**.
The (OAc/TotalOAc)/(NOAc/TotalNOAc) ratio (across all disciplines and years) increases as citation count (c) increases (r = .98, N=6, p<.005). The more cited an article, the more likely that it is OA. (Hajjem et al. 2005)

The question, then, is about *causality*: Are self-archived articles more likely to be cited because they are self-archived (QA)? Or are articles more likely to be self-archived because they are more likely to be cited (QB)?

The most likely answer is that both factors, QA and QB, contribute to the OAA: the higher quality papers gain more from being made more accessible (QA: indeed the top 10% of articles tend to get 90% of the citations). But the higher quality papers are also more likely to be self-archived (QB).

As we will see, however, the evidence to date, because it has been based exclusively on self-selected (voluntary) self-archiving, is equally compatible with (i) an exclusive QA interpretation, (ii) an exclusive QB interpretation or (iii) the joint explanation that is probably the correct one.

The only way to estimate the independent contributions of QA and QB is to compare the OAA for *self-selected* (voluntary) self-archiving with the OAA for *imposed* (obligatory) self-archiving. We report some preliminary results for this comparison here, based on the (still small sample of) Institutional Repositories that already have self-archiving mandates (chiefly CERN, U. Southampton, QUT, U. Minho, and U. Tasmania).

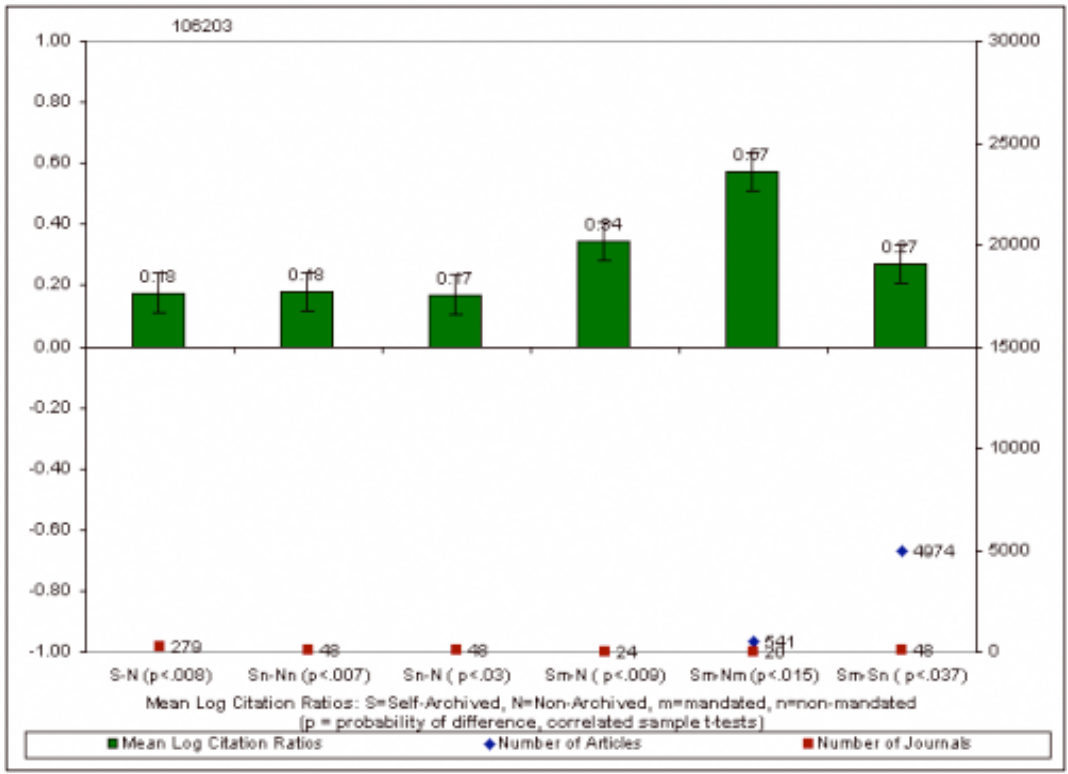

**FIGURE 3. Self-Selected Self-Archiving vs. Mandated Self-Archiving: Within-Journal Citation Ratios** (for 2004, all fields).
**S** = citation counts for articles self-archived at institutions with (**Sm**) and without (**Sn**) a self-archiving mandate. **N** = citation counts for non-archived articles at institutions with (**Nm**) and without (**Nn**) mandate (i.e., **Nm** = articles not yet compliant with mandate). Grand average of (log) **S/N** ratios (106,203 articles; 279 journals) is the OA advantage (18%); this is about the same as for **Sn/Nn** (27972 articles, 48 journals, 18%) and **Sn/N** (17%); ratio is higher for **Sm/N** (34%), higher still for **Sm/Nm** (57%, 541 articles, 20 journals); and **Sm/Sn** = 27%, so self-selected self-archiving does not yield more citations than mandated; rather the reverse. (All six within-pair differences are significant: correlated sample t-tests.) (NB: preliminary, unrefereed results.)

**Summary:** These preliminary results suggest that *both* QA and QB contribute to OAA, and that the contribution of QA is greater than that of QB.

**Discussion:** On Fri, 8 Dec 2006, Henk Moed [**HM**] wrote:

> **HM:** "*Below follow some replies to your comments on my preprint 'The effect of 'Open Access' upon citation impact: An analysis of ArXiv's Condensed Matter Section*'...
>
> "**1. Early view effect. [EA]** *In my case study on 6 journals in the field of condensed matter physics, I concluded that the observed differences between the citation age distributions of deposited and non-deposited ArXiv papers can to a large extent - though not fully - be explained by the publication delay of about six months of non-deposited articles compared to*

> *papers deposited in ArXiv. This outcome provides evidence for an early view [**EA**] effect upon citation impact rates, and consequently upon ArXiv citation impact differentials (CID, my term) or Arxiv Advantage (AA, your term)."*

> **SH:** "The basic question is this: Once the AA (Arxiv Advantage) has been adjusted for the "head-start" component of the EA (by comparing articles of equal age -- the age of Arxived articles being based on the date of deposit of the preprint rather than the date of publication of the postprint), how big is that adjusted AA, at each article age? For that is the AA without any head-start. Kurtz never thought the EA component was merely a head start, however, for the AA persists and keeps growing, and is present in cumulative citation counts for articles at every age since Arxiving began".

> **HM:** "*Figure 2 in the interesting paper by Kurtz et al. (IPM, v. 41, p. 1395-1402, 2005) does indeed show an increase in the very short term average citation impact (my terminology; citations were counted during the first 5 months after publication date) of papers as a function of their publication date as from 1996. My interpretation of this figure is that it clearly shows that the principal component of the early view effect is the head-start: it reveals that the share of astronomy papers deposited in ArXiv (and other preprint servers) increased over time. More and more papers became available at the date of their submission to a journal, rather than on their formal publication date. I therefore conclude that their findings for astronomy are fully consistent with my outcomes for journals in the field of condensed matter physics.*"

The findings are definitely consistent for Astronomy and for Condensed Matter Physics. In both cases, most of the observed OAA came from the self-archiving of preprints before publication (EA).

Moreover, in Astronomy there is already 100% "OA" to all articles after publication, and this has been the case for years now (for the reasons Michael Kurtz and Peter Boyce have pointed out: all research-active astronomers have licensed access as well as free ADS access to all of the closed circle of core Astronomy journals: otherwise they simply cannot be research-active). This means that there is only room for EA in Astronomy's OAA. And that means that in Astronomy all the questions about QA vs QB (self-selection bias) apply only to the self-archiving of prepublication preprints, not to postpublication postprints, which are all effectively "OA."

To a lesser extent, something similar is true in Condensed-Matter Physics (CondMP): In general, research-active physicists have better access to their required journals via online licensing than other fields do (though one does wonder about the "non-research-active" physicists, and what they could/would do if they too had OA!). And CondMP too is a preprint self-archiving field, with most of the OAA differential again concentrated on the prepublication preprints (EA). Moreover, Moed's test for whether or not a paper was self-archived was based entirely on its presence/absence in ArXiv (as opposed to elsewhere on the Web, e.g., on the author's website or in the author's Institutional Repository).

Hence Astronomy and CondMP are fields that are "biassed" toward EA effects. It is not surprising, therefore, that the lion's share of the OAA turns out to be EA in these fields. It also means that the remaining variance available for testing QA vs. QB in these fields is much narrower than in fields that do not self-archive preprints only, or mostly.

Hence there is no disagreement (or surprise) about the fact that most of the OAA in Astronomy and CondMP is due to EA. (Less so in the slower-moving field of maths; see: "Early Citation Advantage?".)

> **SH:** "The fact that highly-cited articles (Kurtz) and articles by highly-cited authors (Moed) are more likely to be Arxived certainly does not settle the question of cause and effect: It is just as likely that better articles benefit more from Arxiving (QA) as that better authors/articles tend to Arxive/be-Arxived more (QB)."

> **HM: "2. Quality bias.** *I am fully aware that in this research context one cannot assess whether authors publish [sic] their better papers in the ArXiv merely on the basis of comparing citation rates of archived and non-archived papers, and I mention this in my paper. Citation rates may be influenced both by the 'quality' of the papers and by the access modality (deposited versus non-deposited). This is why I estimated author prominence on the basis of the citation impact of their non-archived articles only. But even then I found evidence that prominent, influential authors (in the above sense) are overrepresented in papers deposited in ArXiv.*"

I agree with all this: The probable quality of the article was estimated from the probable quality of the author, based on citations for non-OA articles. Now, although this correlation, too, goes both ways (are authors' non-OA articles more cited because their authors self-archive more or do they self-archive more because they are more cited?), I do agree that the correlation between self-archiving-counts and citation-counts for non-self-archived articles by the same author is more likely to be a QB effect. The question then, of course, is: What proportion of the OAA does this component account for?

> **HM:** "*But I did more that that. I calculated Arxiv Citation Impact Differentials (CID, my term, or ArXiv Advantage, AA, your term) at the level of individual authors. Next, I calculated the median CID over authors publishing in a journal. How then do you explain my empirical finding that for some authors the citation impact differential (CID) or ArXiv Advantage is positive, for others it is negative, while the median CID over authors does not significantly differ from zero (according to a Sign test) for all journals studied in detail except Physical Review B, for which it is only 5 per cent? If there is a genuine 'OA advantage' at stake, why then does it for instance not lead to a significantly positive median CID over authors? Therefore, my conclusion is that, controlling for quality bias and early view effect, in the sample of 6 journals analysed in detail in my study, there is no sign of a general 'open access advantage' of papers deposited in ArXiv's Condensed Matter Section.*"

My interpretation is that EA is the largest contributor to the OAA in this preprint-intensive field (i.e., most of the OAA comes from the prepublication component) and that there is considerable variability in the size of the (small) residual (non-EA) OAA. For a small sample, at the individual journal level, there is not enough variance left for a significant OAA, once one removes the QB component too. Perhaps this is all that Henk Moed wished to imply. But the bigger question for OA concerns all fields, not just those few that are preprint-intensive and that are relatively well-heeled for access to the published version. Indeed, the fundamental OA and OAA questions concern the postprint (not the preprint) and the many disciplines that *do* have access problems, not the happy few that do not!

The way to test the presence and size of both QB and QA in these non-EA fields is to *impose* the OA, preferably randomly, on half the sample, and then compare the size of the OAA for imposed ("mandated") self-archiving (Sm) with the size of the OAA for self-selected ("nonmandated") self-archiving (Sn), in particular by comparing their respective ratios to non-self-archived articles in the same journal and year: Sm/N vs. Sn/N).

If Sn/N > Sm/N then QB > QA, and vice versa. If Sn/N = 1, then QB is 0. And if Sm/N = 1 then QA is 0.

It is a first approximation to this comparison that has just been done (**FIGURE 3**) by my doctoral student, Chawki Hajjem, across fields, for self-archived articles in five Institutional Repositories (IRs) that have OA self-archiving mandates, for 106,203 articles published in 276 biomedical journal 2004, above.

The mandates are still very young and few, hence the sample is still small; and there are many potential artifacts, including selective noncompliance with the mandate as well as disciplinary bias. But the preliminary results so far suggest that (1) QA is indeed > 0, and (2) QA > QB.

[I am sure that we will now have a second round from die-hards who will want to argue for a selective-compliance effect, as a 2nd-order last gasp for the QB-only hypothesis, but of course that loses all credibility as IRs approach 100% compliance: We are analyzing our mandated IRs separately now, to see whether we can detect any trends correlated with an IR's %OA. But (except for the die-hards, who will never die), I think even this early sample already shows that the OA advantage is unlikely to be only or mostly a QB effect.]

> **HM: "3. Productive versus less productive authors.** *My analysis of differences in Citation Impact differentials between productive and less productive authors may seem "a little complicated". My point is that if one selects from a set of papers deposited in ArXiv a paper authored by a junior (or less productive) scientist, the probability that this paper is co-authored by a senior (or more productive) author is higher than it is for a paper authored by a junior scientist but not deposited in ArXiv. Next, I found that papers co-authored by both productive and less productive authors tend to have a higher citation impact than articles authored solely by less productive authors, regardless of whether these papers were deposited in ArXiv or not. These outcomes lead me to the conclusion that the observed higher CID for less productive authors compared to that of productive authors can be interpreted as a quality bias."*

It still sounds a bit complicated, but I think what you mean is that (1) mixed multi-author papers (ML, with M = More productive authors, L = less productive authors) are more likely to be cited than unmixed multi-author (LL) papers with the same number of authors, and that (2) such ML papers are also more likely to be self-archived. (Presumably MM papers are the most cited and most self-archived of multi-author papers.)

That still sounds to me like a variant on the citation/self-archiving correlation, and hence intepretable as either QA or QB or both. (Chawki Hajjem has also found that citation counts are positively correlated with the number of authors an article has: this could either be a self-citation bias or evidence that multi-authored paper tend to be better ones.)

> **HM: "4. General comments.** *In the citation analysis by Kurtz et al. (2005), both the citation and target universe contain a set of 7 core journals in astronomy. They explain their finding of no apparent OA effect in his study of these journals by postulating that "essentially all* *astronomers have access to the core journals through existing channels". In my study the target set consists of a limited number of core journals in condensed matter physics, but the citation universe is as large as the total Web of Science database, including also a number of more peripherical journals in the field. Therefore, my result is stronger than that obtained by Kurtz at al.: even in this much wider citation universe, I do not find evidence for an OA advantage effect."*

I agree that CondMP is less preprint-intensive, less accessible and less endogamous than Astrophysics, but it is still a good deal more preprint-intensive and accessible than most fields (and I don't yet know what role the exogamy/enodgamy factor plays in either citations or the OAA: it will be interesting to study, among many other candidate [metrics](#), once the entire literature is OA).

> **HM:** "*I realize that my study is a case study, examining in detail 6 journals in one subfield. I fully agree with your warning that one should be cautious in generalizing conclusions from case studies, and that results for other fields may be different. But it is certainly not an unimportant case. It relates to a subfield in physics, a discipline that your pioneering and stimulating work ([Harnad and Brody, D-Lib Mag., June 2004](#)) has analysed as well at a more aggregate level. I hope that more case studies will be carried out in the near future, applying the methodologies I proposed in my paper.*"

Your case study is very timely and useful. However, [robot-based studies](#) based on much [larger samples](#) of journals and articles have now confirmed the OAA in many more fields, most of them not preprint-based at all, and with access problems more severe than those of physics.

## Conclusions

I would like to conclude with a summary of the "QB vs. QA" evidence to date, as I understand it:

(1) [Many studies](#) have reported the OA Advantage, across many fields.

(2) Three studies have reported QB in preprint-intensive fields that have either no postprint access problem or markedly less than other fields (astrophysics, condensed matter, mathematics).

(3) The author of one of these three studies is pro-OA ([Kurtz](#), who is also the one who drew my attention to the QA [counterevidence](#)); the author of the second is neutral ([Moed](#)); and the author of the third might (I think -- I'm not sure) be mildly anti-OA (Davis -- now collaborating with a publisher to do a 4-year [sic!] [long-term study](#) on QA vs QB).

> Henneken, E. A., Kurtz, M. J., Eichhorn, G., Accomazzi, A., Grant, C., Thompson, D., and Murray, S. S. (2006) [Effect of E-printing on Citation Rates in Astronomy and Physics](#). *Journal of Electronic Publishing*, Vol. 9, No. 2, Summer 2006
>
> Moed, H. F. (2006, preprint) [The effect of 'Open Access' upon citation impact: An analysis of ArXiv's Condensed Matter Section](#)
>
> Davis, P. M. and Fromerth, M. J. (2007) [Does the arXiv lead to higher citations and reduced publisher downloads for mathematics articles](#)? *Scientometics*,

> accepted for publication. *See critiques:* **1**, **2**
>
> (4) *So the overall research motivation for testing QB is not an anti-OA motivation.*
>
> (5) On the other hand, the motivation on the part of some publishers to put a strong self-serving spin on these three QB findings is of course very anti-OA and especially, now, anti-OA-self-archiving-mandate. (That's quite understandable, and no problem at all.)
>
> (6) In contrast to the three studies that have reported what they interpret as evidence of QB (Kurtz in astro, Moed in cond-mat and Davis in maths), there are the many other studies that report large OA citation (and download) advantages, across a large number of fields. Those who have interests that conflict with OA and OA self-archiving mandates are ignoring or discounting this large body of studies, and instead just spinning the three QB reports as their justification for ignoring the larger body of findings.

This will all be resolved soon, and the outcome of our QA vs. QB comparison for mandated vs. self-selected self-archiving already heralds this resolution. I am pretty confident that the empirical facts will turn out to have been the following: Yes, there is a QB component in the OA advantage (especially in the preprinting fields, such as astro, cond-mat and maths). But that QB component is neither the *sole* factor nor the *largest* factor in the OA advantage, particularly in the non-preprint fields with access problems -- and those fields constitute the vast majority. That will be the outcome that is demonstrated, and eventually not only the friends of OA but the foes of OA will have no choice but to acknowledge the new reality of OA, its benefits to research and researchers, and its immediate reachability through the prompt universal adoption of OA self-archiving mandates.

**Stevan Harnad & Chawki Hajjem**
American Scientist Open Access Forum